\documentclass[twocolumn,twoside,slac_two]{revtex4}
\usepackage{graphicx}
\usepackage{fancyhdr}
\usepackage{float}
\begin{document}

\title{Multi-wavelength Observations of PKS 2142-758 during an Active Gamma-Ray State}

%

\author{Michael Dutka}
\affiliation{The Catholic University of America, 620 Michigan Ave., N.E.  Washington, DC 20064}
\author{Roopesh Ojha}
\affiliation{NASA Goddard Space Flight Center Astrophysics Science Division, Code 661, Greenbelt, MD 20771}
\author{Katja Pottschmidt}
\affiliation{Center for Research and Exploration in Space Science and Technology (CRESST) and NASA Goddard Space Flight Center
Astrophysics Science Division, Code 661
Greenbelt, MD 20771 \& Center for Space Science and Technology (CSST)
University of Maryland Baltimore County
1000 Hilltop Circle,
Baltimore, MD 21250}
\author{Justin Finke}
\affiliation{Naval Research Laboratory
Space Science Division, Code 7653,
4555 Overlook Ave. SW,
Washington, DC 20375}
\author{Jamie Stevens}
\affiliation{CSIRO Astronomy and Space Science, Locked Bag 194, Narrabri NSW 2390, Australia}
\author{Jay Blanchard}
\affiliation{University of Tasmania, Newnham Dr, Newnham TAS 7248, Australia}
\author{Roberto Nesci}
\affiliation{University La Sapienza, La Sapienza - Citt� Universitaria, 00161 Rome, Italy}
\author{Philip Edwards}
\affiliation{CSIRO Astronomy and Space Science, PO Box 76, Epping NSW 1710, Australia}
\author{Jim Lovell}
\affiliation{University of Tasmania, Newnham Dr, Newnham TAS 7248, Australia }
\author{Matthias Kadler}
\affiliation{Lehrstuhl f\"ur Astronomie, Universit\"at W\"urzburg, Campus Hubland Nord, Emil-Fischer-Stra�e 31, D-97074 W\"urzburg, Germany}
\author{Joern Wilms}
\affiliation{Remeis Observatory \& ECAP
Sternwartstr. 7, 96049 Bamberg, Germany}
\author{Gino Tosti}
\affiliation{University of Perugia Piazza Universit�, 1, 06123 Perugia, Italy}
\author{Tapio Pursimo}
\affiliation{Nordic Optical Telescope
Apartado 474
E-38700 Santa Cruz de La Palma
Santa Cruz de Tenerife, Spain}

\author{on behalf of the {\em Fermi}/LAT and TANAMI Collaborations}

\begin{abstract}

PKS 2142$-$758 is a flat spectrum radio quasar which emits few, weak but
significant $\gamma$-ray flares in the MeV through GeV energy range.  The
first flare occured on April 4th, 2010, when the source reached a daily flux of
$(1.1 \pm 0.3) \times 10^{-6} $ph cm$^{-2}$ s$^{-1}$ (ATEL \#2539) in
the 100\,MeV to 300\,GeV range. This flux represented more than an order
of magnitude increase over its quiescent flux.  Since the initial
flare, this source has been detected in an elevated state within the
same energy range from October to November of 2010 and another period
ranging from July to August of 2011.  During the latest flaring period
in 2011 a multi wavelength observing campaign was carried out using
the Ceduna radio telescope, the Australian Telescope Compact Array (ATCA),
the TANAMI VLBI Array, {\em Swift}, the Rapid Eye Mount Telescope
(REM), and the Large Area Telescope (LAT) on board {\em Fermi}. These
quasi-simultaneous data were used to construct a broadband SED of this
object in its rare active state. We present these observations and the
resulting SED and some preliminary analysis of the constraints they
place on the high energy emission from this object.

\end{abstract}

\maketitle

\thispagestyle{fancy}


\section{Introduction}
Active galactic nuclei (AGN) are the brightest persistent sources of electromagnetic radiation, from low energy radio waves to high energy $\gamma$-rays.  It is believed that AGN are powered by accretion of matter onto a supermassive black hole \citep{Lynden-Bell1969}.  AGN can produce large scale jets of plasma that can extend for thousands of parsecs.  Blazars are the most luminous and violently variable subclass of AGN whose relativistic jets are believed to be pointed close to our line of sight \citep{Urry1995}.  Blazars are also the most common type of AGN to be detected at GeV energies.  The method by which blazars produce $\gamma$-rays is uncertain and there are a number of  different possible explanations.

\begin{figure*}[t]
\centering
\includegraphics[width=170mm]{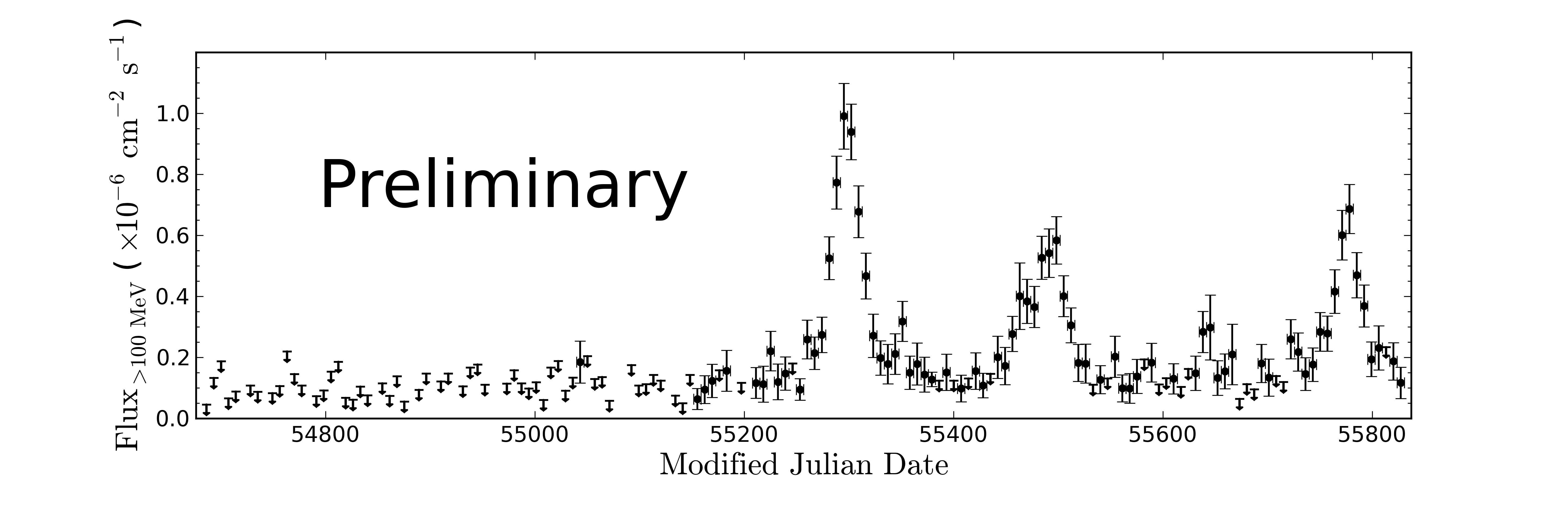}
\caption{$\gamma$-ray light curve of PKS 2142-758 1 week time bins and test statistic $<$ 9 for upper limit threshold}
\end{figure*}
Models of high energy blazar emission can be divided into two major subgroups, leptonic and hadronic models \citep{Bottcher2007}.  Leptonic models focus on a source of seed photons being Compton upscattered to higher energy levels by relativistic electrons in the jet.  Leptonic models can be further subdivided by the source of seed photons, one candidate source is synchrotron photons produced by jet electrons themselves (synchrotron self Compton scattering).  Another possibility is that the seed photons come from a source external to the jet (external Compton scattering).  The accretion disk itself, the broad line region and the dusty torus are all possible examples of photon sources for external Compton scattering.  Seed photon sources do not necessarily have to be confined to the host galaxy, cosmic microwave background photons can also be up scattered to higher energies \citep{Bottcher2008}.  Hadronic models assume that a significant fraction of the jet power is converted into acceleration of protons.  These protons reach the threshold for proton pion production and synchrotron supported pair cascades will develop \citep{Mannheim1992}.   

Quasi simultaneous data on blazar emission at multiple wavelengths across the electromagnetic spectrum will allow us to distinguish between the different models of high energy production within blazars \citep{Bottcher2007}.  The launch of the {\em Fermi} $\gamma$-ray space telescope has created new opportunities for studying blazars at high energies.  The LAT is more sensitive and has better energy and spatial resolution than any previous telescope in the MeV--GeV energy range.  {\em Fermi} has put better constraints on blazar variability within its observing window and has shown that it is not unusual for blazars to exhibit fast variability on timescales of days or even hours at GeV energies.  {\em Fermi} operates in an all sky scanning mode which makes it an ideal rallying point for staging multi instrument observing campaigns on sources in response to their changing $\gamma$-ray state.    

The brightest AGN in {\em Fermi}'s observing window are studied in detail using a broad band multi wavelength approach (e.g.\ \citep{Abdo2010b}, \citep{Pacciani2010}).  However relatively little attention has been given to blazars that are not the most extreme $\gamma$-ray emitters.  PKS 2142-758  is a flat spectrum radio quasar at a redshift of 1.139 \citep{Jauncey1978} and it is not among the brightest {\em Fermi} detected extragalactic sources.  Typical $\gamma$-ray flares of this source reach a flux of  around $1.0 \times 10^{-6} $ph cm$^{-2}$ s$^{-1}$ in the 100\,MeV--300\,GeV energy range and they happen infrequently.   This object will be part of a broader study of blazars which are not the most extreme $\gamma$-ray sources.  The aim of this study is to investigate why some blazars can be very $\gamma$-ray bright while other are not despite their similarities at other wavelengths.         

\section{$\gamma$-ray Properties}
All the $\gamma$-ray data presented in this paper were recorded by the LAT.  The LAT is a high energy $\gamma$-ray telescope with a wide field of view designed to cover the energy band from 20\,MeV to greater than 300\,GeV.  The LAT primarily operates in an all sky scanning mode sweeping its field of view over the entire sky every three hours.   Likelihood analysis is used in order to determine the gamma properties of any detected source for any time range during the LAT's mission lifetime \citep{Atwood2009}.  All LAT data analysis was unbinned and done using the 09-24-00 version of the science tools and the P7SOURCE\_V6 IRF.

\begin{figure}[t]
\includegraphics[width=90mm]{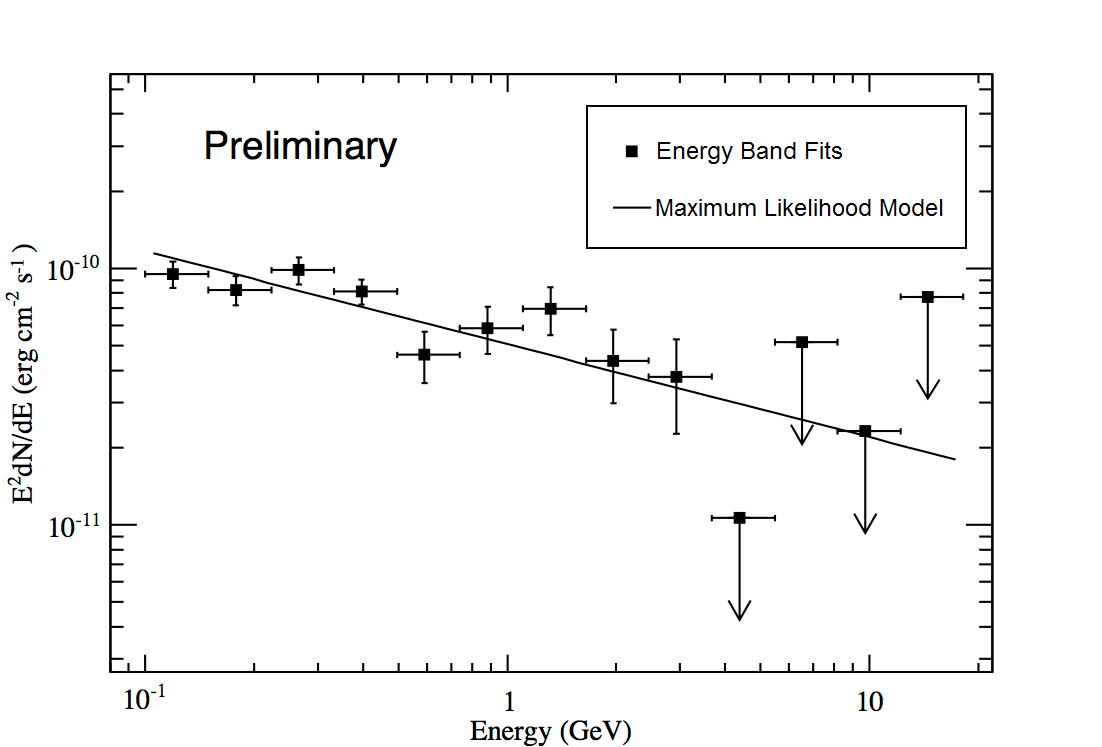}
\caption{LAT energy spectrum, the model displayed is the power law fit for this source across the entire 100\,MeV to 300\,GeV energy range} 
\end{figure}

The light curve (Figure 1) shows the $\gamma$-ray flux of PKS 2142-758 in the 100\,MeV to 300\,GeV energy range averaged into weekly time bins.  A circular region of interest of 10 degrees was used for the analysis.   The light curve begins on August 4, 2008 and ends on September 28, 2011.  Upper limits on the source's flux are calculated when the test statistic (TS) \citep{Mattox1996}  for that week is below 9 (9 roughly corresponds to a 3 sigma detection of the source).  The three flares originating from this source peak at around  $1 \times 10^{-6}$ ph cm$^{-2} $s$^{-1}$ in the 100\,MeV to 300\,GeV range.  Our multi wavelength campaign was carried out during the last flaring period in July and August of 2011. 

The LAT spectrum (Figure 2) was made using a circular ROI of 10 degrees and data from July 21, 2011 to August 18, 2011 which corresponds to a time period when the source's daily TS across the 100\,MeV to 300\,GeV energy band was greater than 25.  Initially a likelihood analysis was run over the whole spectrum to get a power law fit for the $\gamma$-ray emission.  The data were then divided into 13 logarithmic energy bins.  Using the initial power law as a starting point for the analysis the flux was determined in each of the energy bins.  The horizontal bars on each of the data points represent the width of the energy bin and the vertical bars are one sigma errors.  Points  with arrows pointing below them are upper limits.  Upper limits on the target source's flux were calculated when the test statistic within the energy bin was below 25.

\section{{\em Swift} Observations}
{\em Swift} is a multi wavelength space-based observatory that has three different instruments on board, the Burst Alert Telescope (BAT), the Ultra Violet Optical Telescope (UVOT) and the X-Ray Telescope (XRT).  {\em Swift} uses momentum wheels to rapidly change the direction in which it is pointing \citep{Gehrels2004}.  The observatory's primary objective is to observe $\gamma$-ray bursts making its ability to slew rapidly a necessity, however this ability has become extremely useful for studying blazars due to their rapid variability at high energy.  

{\em Swift} observed this source with the UVOT and XRT on August 10, 16, 18, and 23 2011 with exposure times of 2085 s, 4890 s, 3666 s, and 1441 s  respectively.  The most current versions of the {\em Swift} data reduction software (07Jun2011\_V6.11) and calibration data (caldb.indx20110725) were used.  This resulted in four UVOT observations with 3 different filters $U$, $UVW1$ and $UVW2$ which correspond to wavelengths of 3465, 2600 and 1928 \AA\ respectively.  Aperture photometry was done using the standard UVOT data reduction task UVOTSOURCE.   A circular source region of radius 2.5'' and background region of radius 12.5'' were used for the calibration.  UVOTSOURCE also performs coincidence loss corrections on the data.  Magnitudes have been corrected for Galactic extinction using the method described in \cite{Fitzpatrick1999}.

The XRT observations were made in photon counting mode and a 23.6' x 23.6' image is available for each pointing.  The source is clearly visible with no other sources in the region of interest for each pointing.  Identical circular source and background extraction regions of 60" and 119", respectively, were used for all four pointings. Source count rates of 0.054(5) counts/s, 0.040(3) counts/s, 0.049(4) counts/s, and 0.043(6) counts/s were measured.  In order to increase the signal to noise ratios for spectral modeling the spectra of the individual observations were averaged into a single data set.  An overall response file was created weighing the auxiliary response files by photon number and using the appropriate response matrix from the calibration archive, swxpc0to12s6\_20010101v013.rmf. The averaged spectrum was rebinned by combining every 25 channels below 3.5 keV and every 50 channels above. The ~12 ks spectrum was then modeled in the energy range from 0.3\,eV to 8\,keV with an absorbed power law. This resulted in a good fit with best fit parameters of $N_H = 6.6 \times 10^{-20}$ cm$^{-2}$ and Gamma=$1.50^{-0.19}_{+0.20}$ for the neutral hydrogen column density and the power law index, respectively (error confidence level 90\%) and a reduced $\chi^{2}$ value of 1.2 for 10 degrees of freedom.  

\section{Near Infrared Observations with REM}
The Rapid Eye Mount (REM) telescope is a 60\,cm telescope observing at
infrared and optical wavebands. It is able to react quickly to changes
in the sky.  It is located on the La Silla premises of the ESO
Chilean Observatory making it ideal for observing highly variable
southern sources.  Observations of PKS 2142-758 were made on August
10, 2011 shortly after a flare in gamma-ray emission,
with 150\,s exposure time. To calibrate the data, 10 reference stars from 
the 2MASS catalogue were used with J magnitudes ranging from 11.10 to 14.9.  
The linear fit between instrumental and nominal magnitudes was always good, with a slope of
nearly 1, and a dispersion of about 0.07 magnitude. Two near IR flux
measurements of PKS 2142$-$758 were made resulting in magnitudes of
15.05 $\pm$ 0.16 in the $J$ band and 14.92 $\pm$ 0.31 in the $H$ band. The
errors reported are computed by IRAF/apphot and they are based on the
Poisson statistics of the net counts and the detector noise: this
method generally underestimates the actual error in the source's
magnitude. The source was rather faint for REM, so that the total net counts
were rather low and the strongly structured background light leads to an
additional error of 25-30 \% (0.3 magnitude) in the flux
measurement. Our total error for the source flux is about 50\% at
both frequencies.

\section{Radio Observations}
The radio data reported here were obtained from two different
monitoring programs one using the Australia Telescope Compact
Array (ATCA) and the other using the Ceduna telescope.

ATCA has been monitoring TANAMI sources at a number of radio
frequencies.  ATCA data were reduced with the Miriad software package, following standard practices for continuum data. Flux densities were measured using the triple amplitude quantity, and are calibrated against B1934$-$638 for frequencies below 25 GHz, and against Uranus for higher frequencies. One flux density measurement was made for each 2 GHz band, and represents the flux at the centre frequency, assuming that the spectral index across the band does not vary.  For this study, PKS 2142$-$75 was observed on August 30th at 5.5, 17, 19, 38, 40 GHz.

The University of Tasmania operates a 30\,m radio telescope at Ceduna
in South Australia (e.g. \cite{McCulloch2005}). This
telescope has been monitoring TANAMI sources at 6.7\,GHz typically
once every two weeks.  The data are collected in four scan blocks, two
scans in right ascension and two in declination. A block is rejected if any
of its scans fail a Gaussian fit. The data are corrected for gain-elevation effects, 
and a pointing correction applied. B1934$-$638 is used as the primary flux
calibrator.  Variation in system temperature due to ambient temperature 
changes over the course of 24 hours necessitates daily averaging of the data. 
The data used in the construction of this SED were  observed on July 30th, 2011.

\section{SED Modeling}
\begin{figure}[h]
\includegraphics[width=80mm]{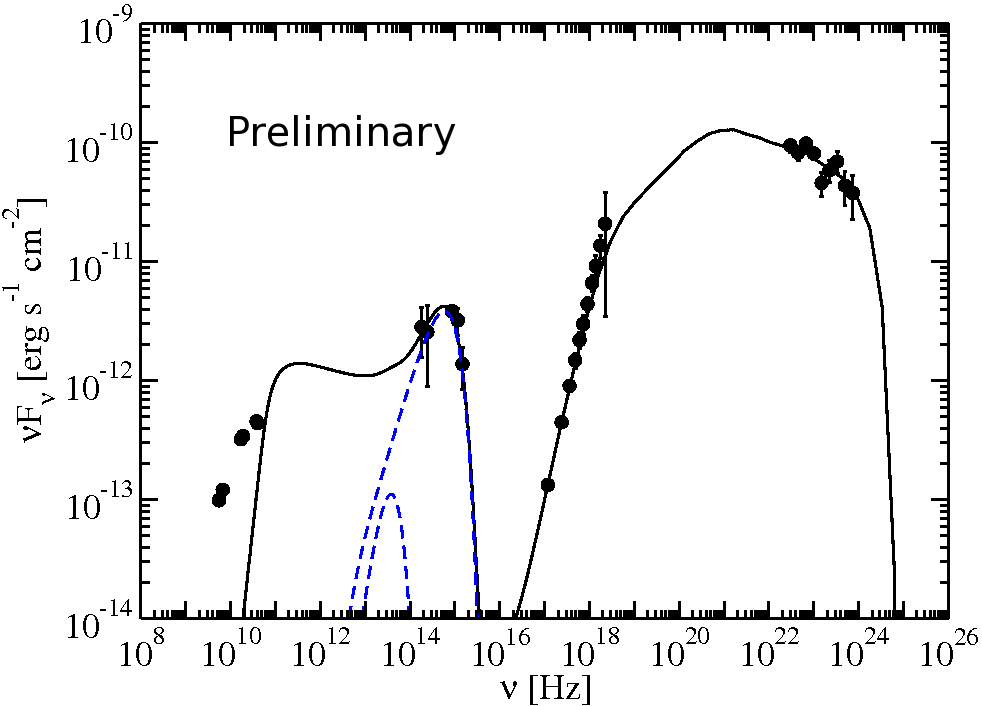}
\caption{Quasi simultaneous SED of PKS 2142-758 while in a $\gamma$-ray active state.} 
\end{figure}

The data from each instrument are plotted on the spectral energy distribution (SED) in Figure 3. An initial attempt at leptonic SED modeling was performed according to the method described in \cite{Dermer2009}.  This model assumes that the radio flux is non-thermal synchrotron radiation emitted by electrons which are isotropically oriented in the co-moving frame of the jet  in a randomly oriented magnetic field in the jet plasma. Optical and UV emission is modeled as thermal emission from the accretion disk.  The X-ray through $\gamma$-ray emission is produced by Compton scattering of seed photons produced in the dust torus.  The contribution of the synchrotron self Compton  effect is very small and plays almost no role, so external Compton processes are the primary source of high energy emission for this source. The variability timescale derived from the LAT light curve constrains the size of the $\gamma$-ray emitting region. The synchrotron self absorption process is included in this fit so the small region that is consistent with the variability timescale cannot explain the radio points.  Table 1 lists the parameters that went into the model.

\begin{table}[h]
\begin{center}
\caption{SED Model Parameters}
\begin{tabular}{|l|c}
\hline \text{Variability Timescale}   \text{$5\times10^{5}$}\,s  \\
\hline Radius of Emission Region = $2.8\times10^{17}$\,cm  \\
\hline Doppler Factor = 40  \\
\hline Magnetic Field = .23\,G \\
\hline \textbf{Broken power law electron distribution:} \\
\hline Minimum electron Lorentz factor = 2.3  \\
\hline Break in Lorentz factor = 230  \\
\hline Maximum electron Lorentz factor = $9\times10^{3}$  \\
\hline Low energy electron spectral index = 2.0  \\
\hline High energy electron spectral index = 3.2  \\
\hline \textbf{Dust torus:}  \\
\hline Dust temperature = $1\times10^{3}$\,K  \\
\hline Dust Luminosity = $2\times10^{45}$\,erg/s  \\
\hline Radius of Dust Torus = 16.2 parsecs  \\
\hline \textbf{Accretion disk emission:}  \\
\hline Black Hole Mass = $1\times10^{9}$ solar masses  \\
\hline Accretion Rate = 1/12 eddington rate  \\
\hline Inner Radius = 6 R$_{g}$ (R$_g$ = $1.5\times10^{14}$ cm)  \\
\hline Outer Radius = $1\times10^{4}$ R$_g$   \\
\hline Luminosity of the disk = $6.5\times10^{45}$ erg/s  \\
\hline \textbf{Equipartition:}  \\
\hline Jet power in magnetic field =  $5.0\times10^{46}$ erg/s  \\ 
\hline Jet power in particles = $2.9\times10^{45}$ erg/s  \\ 
\hline
\end{tabular}
\label{l2ea4-t1}
\end{center}
\end{table}

\section{Conclusions and Future Work}
The next step in this project is to construct and model an SED of the
source when it is in a quiescent state. Comparison of the fitted
parameters in the active and quiescent states should provide
constraints on the gamma-ray production mechanism in this source.
Ongoing VLBI (Very Long Baseline Interferometry) monitoring by the
TANAMI program \citep{Ojha2010} will allow us to
resolve the jet structure of this source and follow its kinematics. In
particular, apparent proper motion measurements will provide a much
stronger constraint on the Doppler factor than is possible with the
SED modeling.

\bigskip 
\begin{acknowledgments}
On behalf of the coauthors I would like to thank the {\em Fermi} and Tanami teams,  Bill McConville for his assistance with {\em Fermi} data reduction, Neil Gehrels and the {\em Swift} team for scheduling the ToO. This research was funded in part by NASA through Fermi Guest
Investigator grants NNH09ZDA001N and NNH10ZDA001N. This research was
supported by an appointment to the NASA Postdoctoral Program at the
Goddard Space Flight Center, administered by Oak Ridge Associated
Universities through a contract with NASA.
\end{acknowledgments}

\bibliographystyle{jwaabib}
\bibliography{aa_abbrv,mnemonic,tanami}





\end{document}